THE EUROPEAN PHYSICAL JOURNAL C

Regular Article - Experimental Physics

# R&D of wavelength-shifting reflectors and characterization of the quantum efficiency of tetraphenyl butadiene and polyethylene naphthalate in liquid argon

G. R. Araujo[1,a] , L. Baudis[1], N. McFadden[1], P. Krause[2], S. Schönert[2], V. H. S. Wu[1]

[1] Department of Physics, Physik-Institut, Universität Zürich, 8057 Zurich, Switzerland
[2] Physik Department, Technische Universität München, 85748 Munich, Germany

**Abstract** Detectors based on liquid argon (LAr) often require surfaces that can shift vacuum ultraviolet (VUV) light and reflect the visible shifted light. For the LAr instrumentation of the LEGEND-200 neutrinoless double beta decay experiment, several square meters of wavelength-shifting reflectors (WLSR) were prepared: the reflector Tetratex® (TTX) was in-situ evaporated with the wavelength shifter tetraphenyl butadiene (TPB). For even larger detectors, TPB evaporation will be more challenging and plastic films of polyethylene naphthalate (PEN) are considered as an option to ease scalability. In this work, we first characterized the absorption (and reflectivity) of PEN, TPB (and TTX) films in response to visible light. We then measured TPB and PEN coupled to TTX in a LAr setup equipped with a VUV sensitive photomultiplier tube. The effective VUV photon yield in the setup was first measured using an absorbing reference sample, and the VUV reflectivity of TTX quantified. The characterization and simulation of the setup along with the measurements and modelling of the optical parameters of TPB, PEN and TTX allowed to estimate the absolute quantum efficiency (QE) of TPB and PEN in LAr (at 87K) for the first time: these were found to be above 67 and 49%, respectively (at 90% CL). These results provide relevant input for the optical simulations of experiments that use TPB in LAr, such as LEGEND-200, and for experiments that plan to use TPB or PEN to shift VUV scintillation light.

## 1 Introduction

LAr is widely used as a scintillating medium in particle physics experiments [1–9]. To facilitate the detection of its VUV scintillation light (peaked at 128 nm [10]), the wavelength shifter TPB is often used [1–8]. TPB absorbs the short-wavelength light and re-emits it in a region between 380 and 600 nm, as shown in Fig. 1 (top). The emission in this spectral region is of advantage: it matches well the quantum efficiency of commercial photodetectors, which often peaks in the violet to blue spectral region [5,11], the efficiency of reflectors, as it will be shown in this work, and the absorption of wavelength-shifting fibers, such as those used in LEGEND-200, shown in Fig. 1 (bottom).

The neutrinoless double beta decay experiment GERDA successfully used TPB to facilitate the detection of scintillation from LAr, which was used to veto events that scatter in both LAr and in the germanium crystals [7,14]. LEGEND-200, which is currently under construction, will use a similar approach [13], as schematically shown in Fig. 2.

To increase the collection of LAr scintillation in LEGEND-200, ∼ 13 m$^2$ of WLSR were prepared and an in-situ vacuum evaporation of TPB on TTX was developed. For the proposed LEGEND-1000 [9], even larger surfaces might require WLSR covering, and the uniform application of TPB becomes more challenging. The scale-up of other experiments that use wavelength shifters (WLS) face similar challenges, which motivates the use of the WLS plastic PEN, as suggested by [15] and thoroughly discussed in [16].

While PEN emits light in a spectral region similar to that of TPB, its advantage over TPB lies in its cost, mechanical stability and ease of manipulation. PEN is commercially available as a film that is flexible at room-temperature (RT) and presents high tensile strength at cryogenic temperatures [17,18]. TPB is a powder that requires specific coating techniques and can emanate in LAr [19]. TPB has been however extensively characterized [19–25], and its QE at 128 nm was measured to be $(60 \pm 4)$% at RT [20]. This value had not yet been measured in LAr, but its efficiency has been observed to increase at low temperatures [22,26].

[a] e-mail: gabriela@physik.uzh.ch (corresponding author)

Springer

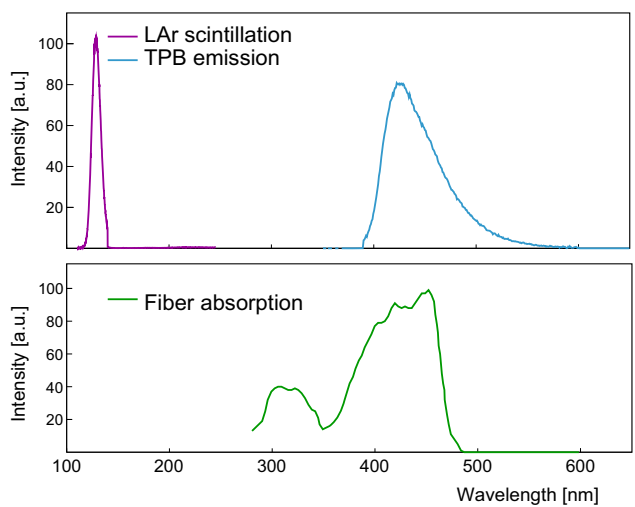

**Fig. 1** Spectra of LAr VUV scintillation [10], TPB emission (top), and absorption of wavelength-shifting fibers used in LEGEND-200 (bottom) [12,13]

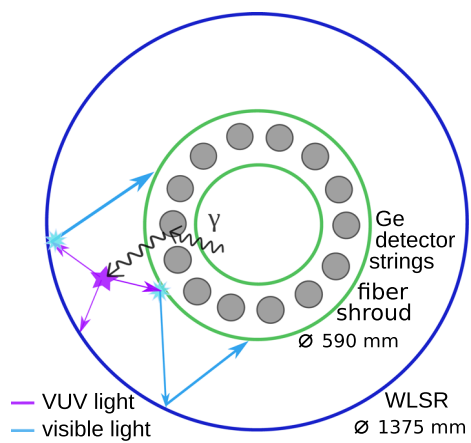

**Fig. 2** In LEGEND-200, the germanium crystals will be immersed in LAr and surrounded by wavelength-shifting fibers. The fibers are coated with TPB, which absorbs the VUV light created by an event in LAr and re-emits it in a spectral region that matches the absorption of the fibers (shown in Fig. 1). The fibers re-emit this light in the green region, which is then detected by silicon photomultipliers. Wavelength-shifting reflectors (WLSR) surround the fibers, shifting the VUV scintillation light and reflecting visible light towards the fibers

The efficiency of PEN has been only measured relative to TPB [15,18,27,28]. The complication of relative measurements is that the light yield from TPB coatings is sample-dependent, with large variations for different coating methods, thickness and substrates [13,20,21,29]. The QE (or quantum yield) represents better the true efficiency, given that it is an intrinsic characteristic of the molecules—namely the ratio between the average number of photons emitted and absorbed [16,20,30]. It is thus not dependent on specific parameters of the sample, such as its thickness, and setup, such as detector position and coverage.

In this work, we investigate combinations of TPB and PEN with reflectors and characterize their absolute QE in LAr for the first time. The goal was to develop and characterize an efficient WLSR for LEGEND-200, as well as for future LAr-based experiments or practical VUV/UV-light detectors.

An efficient PEN or TPB-based WLSR should reflect well the light re-emitted at wavelengths from 360 to 600 nm. As a first step, we thus measured the reflectance of a few substrates and quantified the effective absorption of the TPB and PEN films in response to these wavelengths.

In a second step, the specific TPB-based WLSR from LEGEND-200 and PEN were measured in a LAr setup. This setup was simulated in Geant4 and characterized with an absorbing reference sample, such that the intrinsic QE of TPB and PEN in LAr could be estimated.

## 2 Investigated reflectors and wavelength shifters

The microporous PTFE-based film TTX was acquired from Donaldson with a thickness of 254 $\mu$m [31]. Pieces of TTX were attached to copper plates, which served as mechanical support for the thin reflective film (Fig. 3, left). Given its advantageous mechanical stability and radiopurity, copper foil was also investigated as a reflective substrate for TPB deposition. TPB was then vacuum evaporated onto the surfaces of the copper foil (shown in Fig. 3, right), TTX, and glass substrates. The glass substrates were used in order to obtain TPB samples without reflective backing and for later measurement of the evaporation thickness (performed with a profilometer as in [32]).

PEN was acquired from Goodfellow [33] with a thickness of 125 $\mu$m, biaxial orientation and grade Teonex Q53. The film was sanded with grade P240 sandpaper in random directions, then ultrasonic cleaned in a bath of isopropanol. One piece of PEN was attached (with an air gap in between) to TTX. The samples are listed in Table 1 and the measurements of their reflectance and absorption are described in Sect. 3.

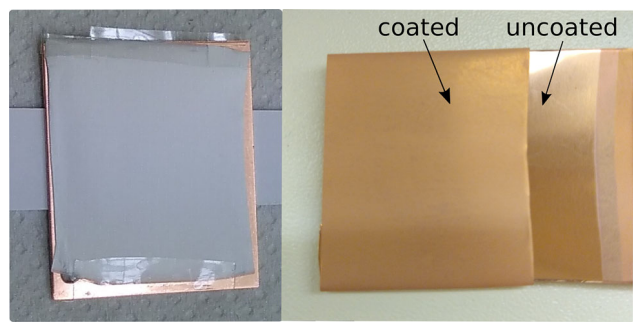

**Fig. 3** A TTX sample attached to a copper plate (left) and a copper foil partially evaporated with TPB (right). The TPB coating changes the appearance of the foil, which becomes slightly white and diffusive

**Table 1** Samples measured with the spectrophotometers (from Sect. 3). The short sample names are shown in the first column. The WLS and reflective films were first measured separately (bare samples). Then, the combinations of WLS films on reflective films (WLSR samples) were measured. All PEN samples are sanded. TPB[(*)] was deposited on thin low-reflective glass, and its measurement is thus considered to be approximately that of 'bare' TPB

| Bare samples | | |
|---|---|---|
| Sample name | Film material | Film thickness |
| TTX | Tetratex® | 254 μm |
| Cu foil | Copper thin foil | 50 μm |
| PEN | Polyethylene naphthalate | 125 μm |
| TPB[(*)] | Tetraphenyl butadiene | 3.0 μm |
| WLSR samples | | |
| Sample name | Reflective film | WLS film | WLS thickness |
| TTX+TPB | TTX | TPB | 0.2, 0.8, 1.0 μm |
| TTX+PEN | TTX | PEN | 125 μm |
| Cu foil + TPB | Cu foil | TPB | 0.8 μm |

**Table 2** Samples prepared for the measurements in LAr (Sect. 4). The short sample names are shown in the first column. All samples were kept in dark after their preparation

| Reflective and WLSR samples | | | |
|---|---|---|---|
| Sample name | Reflective film | WLS film | WLS thickness |
| TTX | TTX | – | – |
| TTX+TPB (L) | TTX | TPB | 0.6 μm |
| TTX+PEN | TTX | PEN | 125 μm |
| Reference sample | | | |
| Sample name | | | Material |
| Absorber | | | Metal Velvet™ foil |

The samples for the measurements in LAr are listed in Table 2. A new sample of TTX+PEN was prepared as previously described. TTX+TPB (L) comes from the evaporation done inside LEGEND-200's cryostat, and is thus a 'witness' of its in-situ evaporated WLSR [13,34]. Special care was taken in this evaporation, as the coating quality is affected by the evaporation conditions (discussed in Appendix A).

To characterize the LAr setup of Sect. 4, the absorbent Metal Velvet™ foil from Acktar [35] was measured as a reference. This foil is well-suited for these measurements as it is manufactured for use at low temperatures, it has low-outgassing properties, and its reflectivity is below 1% from the extreme UV to the infra-red [35].

## 3 Visible light measurements

In this section, we describe the reflectance and absorption measurements of WLSR materials in response to light from 360 to 600 nm, which corresponds to the spectral range emitted by the shifters. As the majority of this light is in the visible spectrum, we will further refer to this range as *vis*.

### 3.1 Setups

For these measurements, we used a LAMBDA 850 UV/vis spectrophotometer from PerkinElmer with a 150 mm integrating sphere (IS). In this sphere, the light reflected (or re-emitted) by the samples is integrated at all angles. As part of the light that comes from samples with TPB or PEN is absorbed by these samples and shifted to longer wavelengths, we used a Cary Eclipse spectrophotometer to resolve the emission wavelengths and decouple these components. As shown schematically in Fig. 4, both setups have monochromators that precisely select the wavelength of light that is directed towards the sample—but while the first provides the integrated angular response (hemispherical reflectivity), only the latter can resolve the wavelength of light that exits the sample.[1]

### 3.2 Procedures, analysis and results

The samples listed in Table 1 were measured in air at room temperature. The setup with the IS was first calibrated using a spectralon standard reflector. The TTX sample was repeatedly measured as a control of systematic uncertainties, which were found to be negligible. Figure 5 shows the integrated intensity measured with the IS.

For the samples that do not shift light, the measured intensity is the absolute reflectance of the sample. For the samples with WLS, this intensity is the sum of the light reflected (or back-scattered) by the sample and the light absorbed by the WLS and then emitted at longer wavelengths. To decouple these contributions, we used the wavelength-resolved measurements shown in Fig. 4b. In these measurements, a total of 12 spectra were acquired for each sample: one per incident

---
[1] To additionally prevent stray light and second order peaks to reach the photodetector in (b), excitation and emission filters were used, which blocked light outside the ranges from 335 to 620 nm and from 360 to 600 nm, respectively.

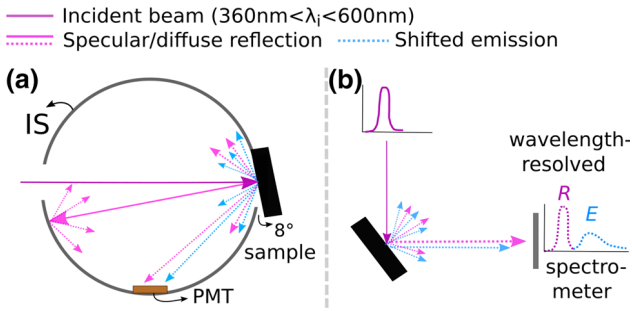

**Fig. 4** **a** In the integrated measurement, the light enters the IS and reaches the sample, which is at an angle of 8°. Most of the diffuse and specular reflection of the sample are integrated by the IS before detection by a Hamamatsu R955 photomultiplier (PMT) [36]. **b** In the wavelength-resolved measurement, the spectra of the reflected/scattered light (R) and shifted emission (E) are measured by a calibrated spectrometer. In both setups, the sample holder backing the sample was not reflective

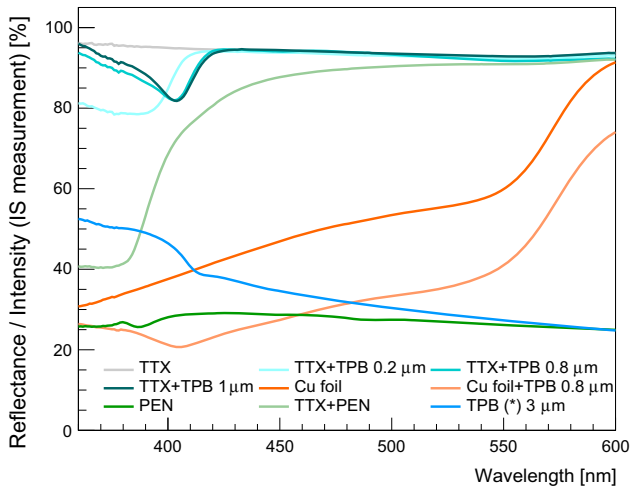

**Fig. 5** Absolute reflectance (or intensity) [%] measured with the IS wavelength ($\lambda_i$) from 370 to 590 nm, with a 20 nm step.[2] An example is shown in Fig. 6, top.

We then integrated the reflection $R(\lambda_i)$ and emission $E(\lambda_i)$ components of each $\lambda_i$-spectrum and calculated the fraction due to the emission component:

$$E_f(\lambda_i) = \frac{E(\lambda_i)}{E(\lambda_i) + R(\lambda_i)} \quad (1)$$

This fraction is used to correct the integrated intensity $I_{IS}(\lambda_i)$ of WLS samples measured with the IS as shown in Fig. 6 (bottom) and described by the following equation:

$$I_{IS}^{corr}(\lambda_i) = I_{IS}(\lambda_i) \cdot [1 - E_f(\lambda_i) \cdot \mathcal{F}(\lambda_i)] \quad (2)$$

---

[2] The entire procedure was performed at 30° and 45°: no significant dependence on the angles of incidence was observed.

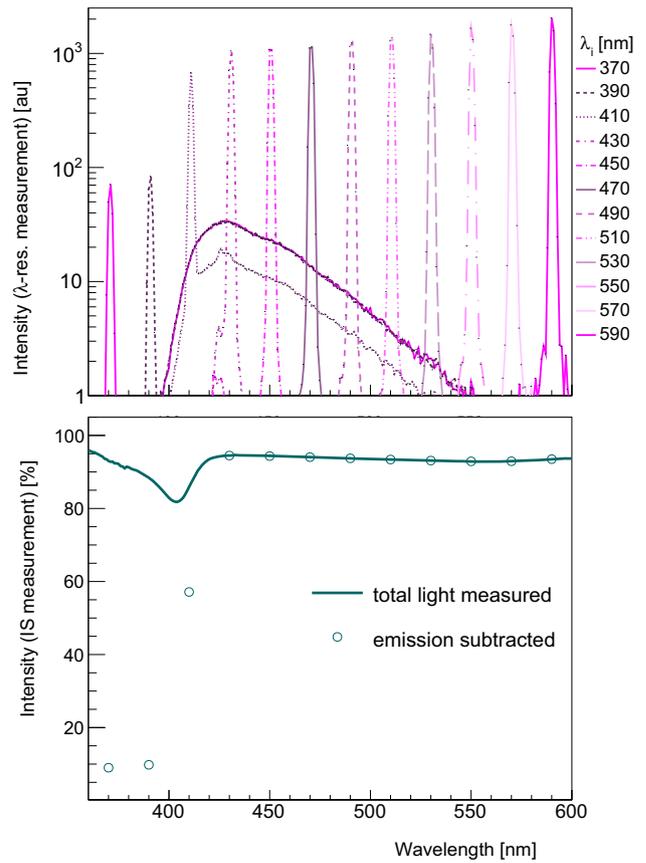

**Fig. 6** Measurements of a TTX+TPB sample. Top: wavelength-resolved spectra for each $\lambda_i$. The peak of reflected light $R(\lambda_i)$ is at the position of $\lambda_i$, and for $\lambda_i \lesssim 410$ nm, a broad peak of shifted emission $E(\lambda_i)$ is also present. Bottom: integrated intensity $I_{IS}(\lambda_i)$ before (—) and after (○) correction of the emission fraction

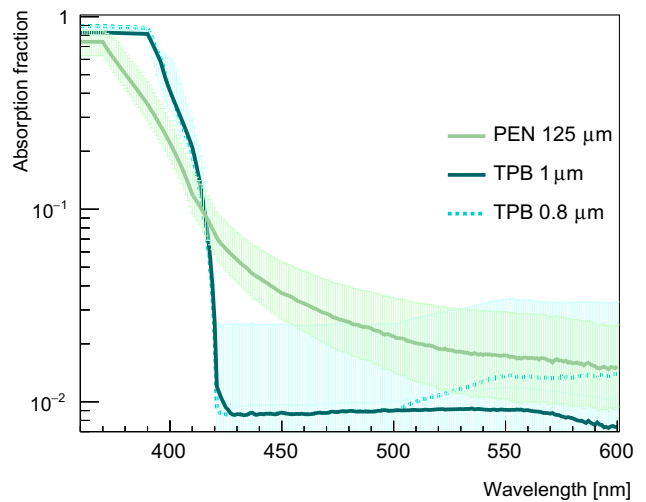

**Fig. 7** Absorption fraction of the PEN and TPB films. The line from TPB 0.8 μm is only clearly visible above 500 nm (the lines almost overlay below this value). The uncertainty bands are discussed in the text

where $\mathcal{F}(\lambda_i)$ is a correction for the PMT response in the wavelength regions of the two components [36]. For the TTX+WLS samples, $I_{IS}^{corr}(\lambda_i)$ can be written as:

$$I_{IS}^{corr}(\lambda_i) = I_R^W(\lambda_i) + I_{TR}^*(\lambda_i) \quad (3)$$

where $I_R^W(\lambda_i)$ is the portion of light that is directly reflected (or scattered) by the WLS film, which corresponds to the $I_{IS}^{corr}(\lambda_i)$ of the shifters measured without reflective backing. $I_{TR}^*(\lambda_i)$ is a more complex convolution of processes: a fraction of the total light intensity traverses the WLS film (with probability $I_T^W(\lambda_i)$), is reflected by TTX with probability $R_{TTX}(\lambda_i)$ (known from Fig. 5), and then traverses the film again. This can be approximately described as:

$$I_{TR}^*(\lambda_i) \approx I_T^W(\lambda_i) \cdot R_{TTX}(\lambda_i) \cdot I_{T'}^W(\lambda_i) \cdot [1 + I_\Gamma(\lambda_i)] \quad (4)$$

where $I_\Gamma(\lambda_i)$ is a correction factor that accounts for multiple scatters between the film and TTX[3], and we consider $I_{T'}^W \approx I_T^W$. Equations 3 and 4 show that the $I_{IS}^{corr}(\lambda_i)$ of the WLSR combinations is a measure of both their efficiency in reflecting vis light, and the vis transparency of the WLS films, $I_T^W(\lambda_i)$ – which can be thus extracted from these equations. We can then obtain the effective absorption fraction of the PEN and TPB films for the vis wavelengths:

$$I_\alpha^W(\lambda_i) = 1 - I_R^W(\lambda_i) - I_T^W(\lambda_i) \quad (5)$$

The results are shown in Fig. 7. For the absorption fraction of the TPB films, we assumed a ±60% uncertainty on the estimation of $I_R^W(\lambda_i)$ from TPB[(*)] – given that the TPB thickness was larger and its substrate was not completely non-reflective.[4] The resulting errors (shaded bands) are small at wavelengths that are mostly absorbed and large above ∼420 nm, where the absorption fraction is low. For the absorption fraction of PEN, we also assume a conservative ±60% uncertainty on $I_R^W(\lambda_i)$.[5]

### 3.3 Discussion

The reflectance and absorption spectra are important parameters for the design and material selection of an efficient WLSR. The reflectance of TTX – shown in Fig. 5 – is ≳ 94.5% at the emission peak of TPB at ∼ 425 nm. The combination of WLS and reflector that presents best reflectance (and lowest absorption) is TPB on TTX. TTX however requires a material for the mechanical support. Thus, other substrates such as copper were investigated. Coating TPB directly on copper – which is radiopure and mechanically stable – or coupling TTX to PEN would avoid the need of a 3-layered structure (copper plate + TTX + TPB). The reflectance of TPB coatings on Cu foils was however proved to be low: less than half of that from TPB on TTX.

Given the white appearance of thick coatings of TPB, we investigated whether a 3 µm layer of TPB could act as the reflector of visible light. The results from TPB[(*)] 3 µm show that TPB alone is not a good reflector; its reflectance is below 40%.

The integrated intensity from the TPB coatings on TTX was approximately independent of coating thickness for light above 420 nm: the values were almost as intense as the ones from the bare TTX. Thus, thicker (∼ 1 µm) coatings of TPB are preferred for a TTX-based WLSR: the coating is thick enough to fully absorb the VUV light[6] while light is not significantly lost to absorption.

The absorption of light above 420 nm by 1 µm of TPB might be well below 1%, as shown by the lower error bands of Fig. 7: thin TPB films might be thus ∼ absorption free in this region. Below 420 nm, TPB absorbs part of the light and partially re-emits it. The absorption edge of Fig. 7 agrees with those measured by [37] and [24].

While sanded PEN reflected ∼30% of light above 420 nm, TTX+PEN reflected more than 80%: similarly to TPB+TTX, the reflectivity is dominated by the TTX backing. Both PEN and TPB films absorb more than 10% of the light below 410 nm. Around the emission peak of the shifters (∼ 425 nm), the PEN film absorbs significantly more than the thin TPB films. This effect is also seen in Fig. 5: the lower intensity of PEN+TTX (compared to bare TTX, or TTX+TPB) is due to the absorption of photons, either by PEN and/or by both PEN and TTX[7] after multiple reflections (within the PEN film or between PEN and TTX).

While TTX+PEN reflects more vis light than Cu foil + TPB, the performance of a PEN-based WLSR could be improved by finding a better coupling to the reflector, or a better reflector, and/or improving its effective vis absorption (by using a thinner film, for instance). These results show that even if the QE of PEN was the same of TPB's, its light yield would be lower, since more of the shifted light is lost for the 125 µm film. Therefore TPB on TTX with a thickness of 600 nm was chosen for the WLSR in LEGEND-200. To investigate further PEN and estimate its QE at low temperature, we also measured it in LAr, along with the sample from LEGEND-200, as described in the next section.

---

[3] A photon could scatter *n*-times between the WLS film and TTX, adding $[R_{TTX}(\lambda_i) \cdot I_{R'}^W(\lambda_i)]^n$ terms to $I_\Gamma$. We neglect high orders (n>3) of these reflections, as the vis-reflectance of PEN or TPB is not high.

[4] The thin glass substrate might contribute to up ∼ 25% of the $I_R^W$ measured from TPB[(*)] (assuming the reflectance of the glass is < 8%).

[5] Although this $I_R^W$ is indeed from the bare PEN, slight differences of its sanded surface and total internal reflection within the film may lead to uncertainties. We thus assume a conservative uncertainty on it.

[6] The optimum light yield was measured for thickness between 0.5 and 1 µm of TPB on TTX [13].

[7] TTX either absorbs or transmits the 1-$R_{TTX}$ fraction of photons.

## 4 LAr measurements and simulation

In this section, we describe the measurements of the WLSR samples in LAr. The setup is first characterized by using an absorber with known optical parameters. Afterwards, the reflective and WLSR samples are measured. These lead to an enhancement of the detected light in comparison to the absorber. A detailed optical simulation of the setup and sample materials was performed. Uncertainties on the measured values and optical parameters are considered, and limits are extracted for the QE of PEN and TPB in response to scintillation from LAr.

### 4.1 Setup

Figure 8 shows the schematic of the LAr setup: a cryostat, equipped with a cooling system, levelmeter and temperature sensors (described in [29,38,39]); and a sample cell, delimited by a sample holder, an $\alpha$ source, and a VUV-vis sensitive PMT.

The 3" PMT (R11065 model from Hamamatsu modified with a $MgF_2$ window) is sensitive to both VUV and visible light, and suited for LAr temperature. In the *vis-only* measurements, an acrylic filter was used in front of its window to limit the sensitivity of the PMT to visible wavelengths. The relative geometry of the PMT, source, and sample was kept the same for all the measurements.

The $\alpha$ source is $^{241}$Am deposited on a thin stainless steel disc [29]. The source has an activity of 30 Bq, is non-encapsulated, and its main emission are 5.486 MeV $\alpha$'s. This source of scintillation photons is optimal: well localized and with energy higher than most background sources. The source is located at the base of the sample cell, so that the scintillation photons from $\alpha$-decays can either reach the sample or be directly detected by the PMT.

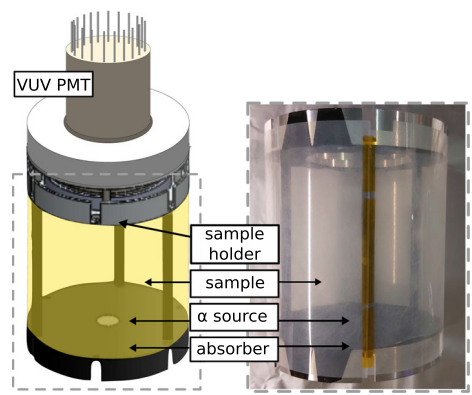

**Fig. 9** Left: schematic of the sample cell. Right: picture of the sample holder covered with the absorber (described in Sect. 2) and surrounded by sanded PEN, which is shown before attaching TTX to it. Kapton tape is used to close its volume from outside (behind the support rod). The PMT limits the sample cell on the top

Apart from a small opening for the source (Ø 7 mm), all structural parts in the sample cell are covered with the low-outgassing absorber from Acktar, as shown in Fig. 9. In this way, uncertainties from the reflectivity of other structural materials are avoided.

### 4.2 Procedures

The samples listed in Table 2 were measured. Each sample had a rectangular size of 110 mm x 350 mm. They were wrapped around the cylindrical sample holder and closed (from outside) with Kapton tape, as shown in Fig. 9. In the case of PEN, first PEN was mounted – as shown in the same figure - and then TTX was wrapped around it. Two ring-shaped aluminum clips were used to hold the sample against the top and bottom parts of the holder.

After mounting a sample, the cryostat was closed and pumped to approximately $10^{-5}$ mbar. Then it was filled with gaseous argon (GAr) with 99.9999% purity, which was liquefied inside the cryostat using the system described in [29,38]. GAr was filled until the LAr level in the cryostat was a few cm above the PMT window. The level could be verified by reading the levelmeter as well as the temperature of the sensor located on the PMT. The setup was always kept in overpressure (above 1 and below 1.3 bar) and the temperature during operation was between 87 and 89 K. The pressure and temperature were monitored continuously.

The PMT was biased at 1350 or 1500 V and set to trigger on single-photoelectrons (SPE), which were later used for the calibration of the signal. Data taking started after the LAr level, temperature and pressure were stable. The signals were acquired without amplification by a CAEN V1724 module with an acquisition rate of 100 MHz, an acquisition window of 8 µs, a pre-trigger of 1 µs and a trigger level at 3 sigma above the baseline. All the waveforms were saved for later processing. Data was taken for the two PMT voltages every day for a period of three to four days. Each data taking setting, for each day and voltage, is considered as one run.

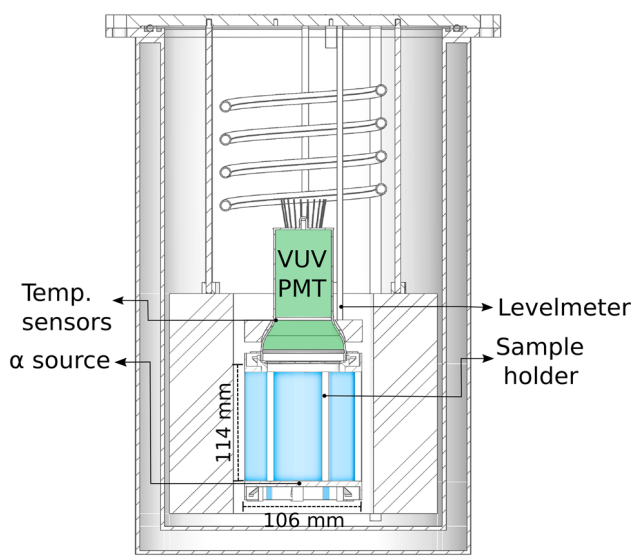

**Fig. 8** Schematic cut of the LAr setup with its main components indicated

The samples were measured in 3 different modes: (i) *VUV-only*: used to characterize the setup and the reflector TTX; (ii) *VUV+vis*: both VUV light that went direct to the PMT as well as light shifted by the WLSR samples are measured; (iii) *vis-only*: VUV-light is blocked by the acrylic filter and only shifted light is measured.

At the end of a sample measurement, the LAr was evaporated and not reused. This was done to prevent the accumulation of impurities in LAr, which can decrease its scintillation yield [40]. To avoid variating the purity level, the material content and vacuum level of the cryostat at the time of filling were always roughly the same. The systematical uncertainty on the effective scintillation yield within a sample measurement was then assessed by taking several runs, at least two days apart, and by repeating the measurement of the WLS samples in the *vis-only* mode. Further assessment on this systematic uncertainty was performed at the analysis level, by investigating the signal waveforms, namely the triplet lifetimes and prompt light emission, as described in the next section.

### 4.3 Data analysis

To determine the number of photoelectrons (PE) detected when 5.486 MeV are deposited by $\alpha$'s in the LAr of the cell, we first integrate the acquired waveforms within the 7 µs of the post-trigger window. The histogram of these integrals present the characteristic SPE peak and the pronounced $\alpha$-peak, as shown in Fig. 10. We fit the SPE peak of each run with a Gaussian function, an then translate the integrals of the signals into PE values. This is used as a calibration of the PMT for each run, ensuring that possible drifts in the voltage and gain of the PMT do not affect the final results.

The mean PE value of the $\alpha$-peak is then fitted for each run, and for three selected datasets: (i) the full dataset, (ii) the dataset after applying pulse-shape discrimination (PSD), and (iii) a *prompt* dataset, for which only the first 200 ns of the waveform is integrated. The histogram of the dataset after PSD is also shown in Fig. 10: the $\alpha$-peak is in the same position, since the PSD cut only reduces the background from electron recoils (ER) while keeping the signal from the $\alpha$ events, as shown in Fig. 11. The low-energy tail of the PE peak present after PSD is due to $\alpha$'s which deposit only part of their energy in LAr. We only use the main peak for estimating the mean PE value.

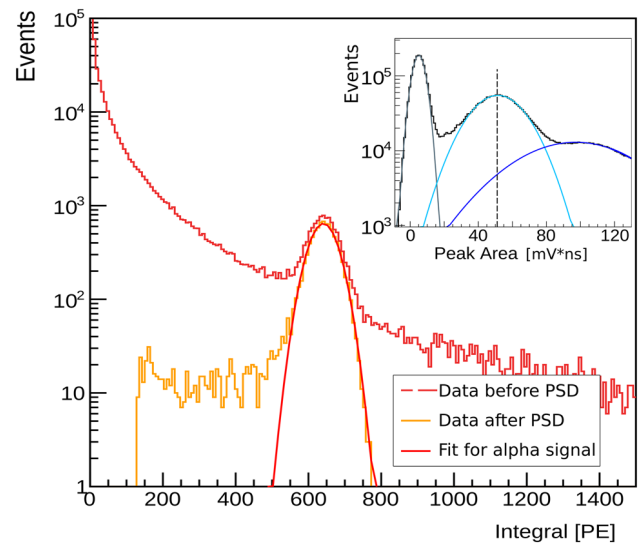

**Fig. 10** Integral histogram of a given run of the TTX sample measurement before and after the pulse shape discrimination. Inset shows a zoom in the SPE peak region, which is used for the PE calibration

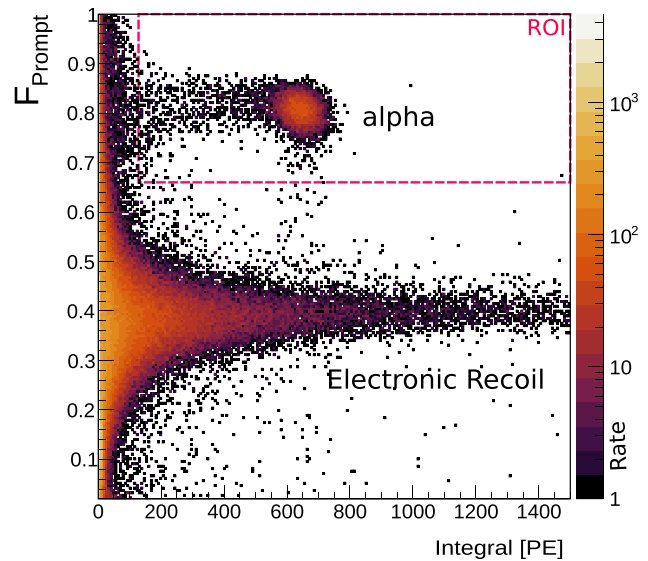

**Fig. 11** Integral of the waveform versus the ratio $F_{Prompt}$ of the light integrated in the first 200 ns to that of the whole waveform, similarly to done in [41]. The alpha signal presents a higher ratio of prompt light, and can be discriminated from the background-induced ER band. The region of interest (ROI) is marked in red

The mean PE value and its uncertainty for each sample measurement is calculated from the distribution of the values obtained from the fits of data from each run. One mean PE value is thus obtained for each sample measurement and data selection. As the values obtained from the datasets before and after PSD agree well, only the values after PSD are shown in Table 3. As only part of the signal is integrated for the *prompt* dataset, this dataset is not used to estimate the light detected, but rather as an investigation as to whether the prompt light

**Table 3** List of measurements and detected PE values for each measurement mode. $\mathcal{R}$ indicates the intensity of light detected relative to the reference absorber measurement. The $\mathcal{R}^{(p)}$ is the same ratio but for the *prompt* analysis. $\tau_t$ is the estimated triplet lifetime

| Sample | Mode | PE value | $\mathcal{R}$ | $\mathcal{R}^{(p)}$ | $\tau_t$ (µs) |
|---|---|---|---|---|---|
| Absorber | *VUV-only* | $567 \pm 24$ | 1 | 1 | 1.11 |
| TTX | *VUV-only* | $610 \pm 25$ | $1.08 \pm 0.06$ | $1.05 \pm 0.05$ | 1.16 |
| TTX+TPB (L) | *VUV+vis* | $1238 \pm 36$ | $2.18 \pm 0.11$ | $2.11 \pm 0.11$ | 1.21 |
|  | *vis-only* | $747 \pm 33$ | $1.32 \pm 0.08$ | $1.27 \pm 0.07$ | 1.32 |
| TTX+PEN | *VUV+vis* | $1071 \pm 40$ | $1.89 \pm 0.11$ | $1.83 \pm 0.08$ | 1.21 |
|  | *vis-only* | $362 \pm 12$ | $0.64 \pm 0.03$ | $0.60 \pm 0.03$ | 1.27 |

– which is less affected by impurities [40] – yields the same relative ($\mathcal{R}$) values of PE obtained when integrating the full waveforms. As shown by $\mathcal{R}$ and $\mathcal{R}^{(p)}$, in Table 3, all the measurements in this work show this agreement, indicating that the full waveforms (and thus the scintillation yield) are not significantly affected by a systematic error of varying purity levels across measurements.

To estimate the triplet lifetime of the scintillation light ($\tau_t$, shown in Table 3), the waveforms of the ER signal are stacked and fit by an exponential function from 450 to 4000 ns. Errors from fitting different runs of a measurement are negligible and thus not shown. No modelling of the afterpulsing of the PMT and decay lifetime of the WLSs are considered in the fit.[8] The estimated $\tau_t$ of the samples without shifters (TTX and absorber) are similar ($\sim 1.1$ µs), but smaller than the ones estimated from the *vis-only* mode of TPB and PEN ($\sim 1.3$ µs). This is expected for TPB, since its delayed emission causes the estimation of $\tau_t$ to be larger than its actual value [41]. A delayed component has not been reported for PEN, but the fact that its estimated $\tau_t$ is similar to TPB's and increases for the *vis-only* measurement may hint at a similar component.

In Fig. 12, the waveforms of the WLSR measurements show the apparent longer $\tau_t$ from the *vis-only* measurement and the broader singlet component of TTX+PEN compared to that of the TTX+TPB sample.

We note that while the triplet lifetime of LAr is often related to its purity level and scintillation yield [40], the measurements in this work are done in different VUV and vis modes – with the re-emission of the WLSs possibly increasing the estimated $\tau_t$ in different levels. We thus cannot easily relate $\tau_t$ to the purity and scintillation yield levels. We can only compare the $\tau_t$ values within each measurement mode, which were found to be very similar. The value of the triplet lifetime of pure argon is $\approx 1.3$ µs, measured without WLS [10,41]. The lower $\tau_t$ measured in this work ($\approx 1.1$ µs) hints at the presence of impurities, likely due to the outgassing of a PTFE filler present in the cryostat, as discussed in [29].

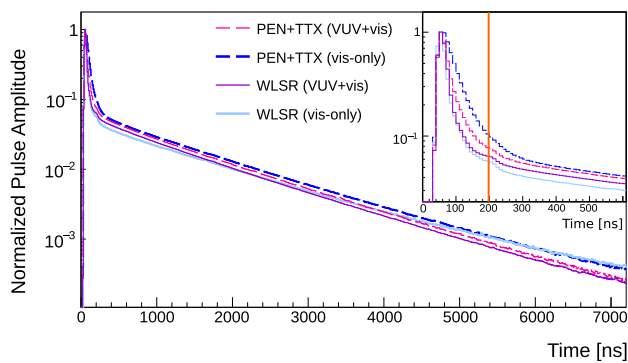

**Fig. 12** Stacked waveforms of different measurements normalized to their peak values. The position defining the prompt signal is marked in orange

The quenching of scintillation yield due to impurities would be a problem if it was significantly unstable, which is not the case, given that: (i) the triplet lifetime of each sample measurement did not systematically vary during a measurement (across runs of a few days) or a given mode, (ii) the *prompt* analysis agrees with the one using the entire waveform, (iii) the repeated measurements of TPB and PEN in the *vis-only* mode lead to consistent results, as it will be shown in the next section.

### 4.4 Simulations and results

In order to extract the intrinsic QE of TPB and PEN, a Monte Carlo simulation of the $\alpha$-decays and scintillation photons in the LAr setup was done using the Geant4 toolkit [42]. The geometry of the sample cell and the optical parameters of its surfaces were implemented in the simulation. To quantify the effective VUV photon yield ($PY_{eff}$) from an $\alpha$-decay in the sample cell – a parameter input in the simulation – we used the absorber measurement. The VUV reflectivity of TTX was then extracted by attributing the increase of detected light in the TTX measurement to its VUV reflectivity. Finally, the QE of TPB and PEN are estimated by considering the PE values measured from the samples in the LAr setup along with the knowledge of $PY_{eff}$, the VUV-vis reflectivity of TTX, and the optical modelling of TPB and PEN. All these steps are described in Sect. 4.4.1 to Sect. 4.4.4, and discussed in Sect. 4.5.

---

[8] As detector effects are not taken into account and the $\alpha$ signal has a lower ratio of the triplet component compared to that of the ER signal, we use only the ER events to estimate $\tau_t$.

*4.4.1 Optical parameters and their uncertainties*

The scintillation spectrum of LAr was taken from [10]. Its scintillation yield ($SY_{LAr}$) is tuned in the simulation in order to match the value measured in the setup covered with the absorber (described in Sect. 4.4.2).

The refractive indices ($n_i$) of LAr are calculated using the Sellmeier coefficients recently measured by [43]. The Rayleigh scattering length is then calculated for each wavelength using the $n_i$-dependent equation from [43], resulting in a length of $\approx 99$ cm at 128 nm. The absorption length of LAr was set to 60 cm in the VUV range[9] and 10 m in the visible range. To investigate the uncertainties on these values, the simulation was run for attenuation lengths up to 30 cm for VUV and 200 cm for visible light: the effects were negligible, given the size of the setup.

The QE of the PMT ($QE_{PMT}$) was set according to the values from its data sheet [47]: 22 and 27% at 128 and 420 nm, respectively. Its value in the VUV is degenerate with $SY_{LAr}$ (detailed in Sect. 4.4.2) – thus no uncertainty was added to it. To investigate the effect of its $MgF_2$ window on the wavelength-dependent QE of the PMT in LAr, the optical parameters of the window were modeled using the refractive indices of $MgF_2$ from [48] and [49][10]. These values yield low reflectance, as expected from $MgF_2$ [49]. The relative transmission of the window could be slightly higher in LAr, showing a greater enhancement in the VUV region [44] – these uncertainties are further discussed in Sect. 4.4.4. Uncertainties on the absolute reflectivity of the window were found to be negligible, especially because the surface directly opposite to the PMT is the absorber, as shown in Fig. 9.

The reflectivity of the absorber was set to its nominal value: 0.7% from the VUV to the visible region. Values of its reflectivity up to 2% lead to negligible increase in the number of PE measured.

Besides the absorber, the only materials viewed by the PMT are the sample and the small source disc. While the vis-reflectivity of the unpolished stainless steel (SS) in the disc is negligible, its VUV reflectivity is relevant, given that the scintillation photons are produced very close to the source.

The VUV reflectivity of SS (at 128 nm) is $(23 \pm 2)$% [50].[11] For the overall VUV reflectivity of the source disc ($SD_{ref}$) we consider that $(70 \pm 15)$% of it is SS (the rest is black source material). This results in $SD_{ref} = (16 \pm 4)$%, a value degenerate with other parameters, such as $QE_{PMT}$. $SD_{ref}$ does not lead to uncertainties in case the reflection is diffuse, only for the specular case (discussed in Sect. 4.4.2).

The reflectance of TTX measured in this work is input in the simulation. As the refractive index of LAr is slightly higher than that of air at visible wavelengths [43], its reflectivity could be slightly lower in LAr. Uncertainties on this value are discussed in Sect. 4.4.4. The VUV reflectivity of TTX is at first not known and is thus measured in LAr: the values obtained are then input in the simulation. The refractive index of TTX is not relevant for this simulation, since the reflectivity is fixed and the surface is set to absorb photons that are not reflected.

The VUV absorption of PEN was set to 100%.[12] For the absorption length ($\lambda_{abs}$) of 128 nm photons by TPB, we consider the value from [20] ($\sim 400$ nm) and lower values, given that the efficiency saturation from TTX+TPB samples in response to VUV was achieved at thickness under 1 μm [13,34] (compared to $\sim 2$ μm measured in [20]). These differences are expected: the packing density, and thus the absorption, of TPB may depend on (i) its morphism – which is related to the coating technique and substrate, and on (ii) the perfect coverage of the sample, discussed in Appendix A. Therein, we analyzed the TPB structure and coverage of the sample with microscopy: the samples were proved to be very homogeneously covered with TPB. Yet, a large range of VUV-$\lambda_{abs}$ are conservatively considered in the simulation: from 250 to 450 nm.

The vis-absorption measurements of the TPB and PEN films were used to estimate the mean values ($M_v$) of their effective vis-$\lambda_{abs}$.[13] Upper and lower ($M_v \pm \delta$) limits on vis-$\lambda_{abs}$ were calculated by propagating the uncertainties accounted for in the measurements with an additional 10% systematic. The resulting upper values ($M_v + \delta$) are input in the simulation such that they provide a conservative lower limit on the QE of the shifters. These conservative vis-$\lambda_{abs}$ limits for wavelengths from 425 to 600 nm are 2 to 10 mm for PEN, and $\sim 1$ mm for TPB (the TPB thin film is thus practically absorption free in this region). Adding scattering

---

[9] While ref. [44] set a lower limit of 110 cm on the attenuation length of 128 nm photons by pure LAr (a value consistent with zero absorption), the values of 66 and 50 cm measured by [45,46] indicate varying absorption lengths for different batches (and impurity levels) of LAr.

[10] [48] provides the room-temperature indices in the visible region and down to 140 nm while [49] provides the values from 200 to 110 nm. Both data agreed in the overlap region (from 140 to 200 nm).

[11] This value was measured from electropolished SS, and is likely higher than that of unpolished SS. We thus take 25% as an upper limit.

[12] As shown in this work, the absorption of PEN continuously increase towards lower wavelengths and is thus likely 100% at VUV.

[13] For this, we use $I_\alpha^W = (1 - I_R^W) \cdot (1 - e^{-d/\lambda_{abs}})$. As there is no knowledge of the exact path of the photon after entering the film, we attribute the absorption to a path length equal to the thickness of the film ($d$). This is an approximation valid for the simulation, as it reproduces the effective absorption observed.

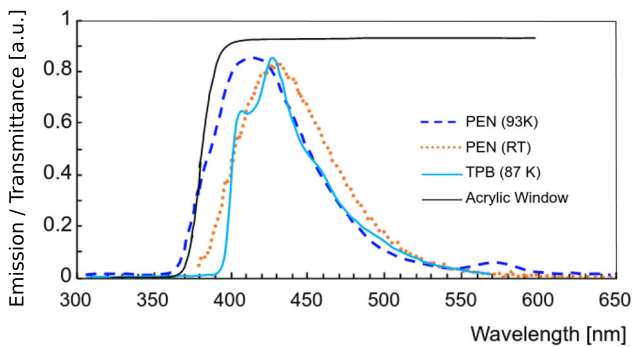

**Fig. 13** Transmittance of the acrylic window and emission spectra of TPB at 87 K [26], of PEN at 93 K [52] and at RT (this work)

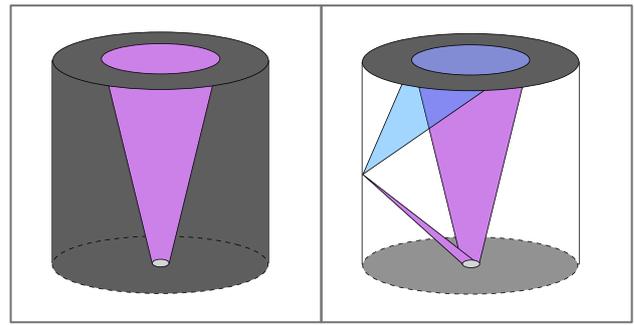

**Fig. 14** Schematic of the light produced by the source in the bottom and detected by the PMT at the top in the *VUV-only* measurement of the absorber (left) and the *VUV+vis* measurement of a WLSR sample (right). For the WLSR, part of the VUV light is absorbed, re-emitted in blue and scattered until it reaches the PMT

lengths above 50 μm for PEN and 2.0 μm for TPB[14] (minimum value from [51]) did not yield significant changes.

The emission spectrum from TPB at 87 K was taken from [26]. The spectrum from PEN was measured at room temperature (RT) with the wavelength-resolved spectrophotometer previously described. We do not use the spectrum from [52], taken at 93 K, because both the RT and 93 K spectra from this reference are slightly shifted towards lower wavelengths when compared to the spectra measured in this work and in [18,32]. While this uncertainty on the spectrum can lead to an error on the QE of PEN estimated from the *vis-only* measurement (some of the spectrum might be cut by the acrylic window, as shown in Fig. 13), the error for the *VUV+vis* measurement is almost negligible – the spectrum used in the simulation yields a slightly more conservative estimation of the QE of PEN.

The refractive indices of TPB and PEN at visible wavelengths were set to 1.62 and 1.51, respectively (values from [37] and [53]). We found that varying these values by ±0.1 for TPB and +0.2 for PEN did not yield significant changes (within uncertainties), given that the reflectance of visible photons by the WLSR samples is dominated by the reflectance of TTX and the absorption of the films, as shown in Sect. 3.3.

The values of these indices at VUV range are not known. Therefore, the surfaces were set to not reflect VUV photons. By turning off the reflection of the VUV photons, one may overestimate the percentage of incident photons that get absorbed. This however allows for conservatively estimating the minimum QE of the shifters.

To input the properties of the poly(methyl methacrylate) (PMMA) acrylic window in the simulation of the *vis-only* mode, we measured its transmittance with a UV–Vis spectrophotometer, shown in Fig. 13. The 92.3% transmittance in the plateau region is consistent with losses due to Fresnel reflection from two surfaces of PMMA with a refractive index of ~1.5 (value from [54]), and negligible absorption. The absorption length of the window was thus set to 1 m in this region. For the values between the plateau and 350 nm, the absorbance was calculated by subtracting the transmittance and the reflectance from one. Below 350 nm, no light is transmitted and all of it is considered absorbed. A cross check of the modelling of the window is done by simulating the absorber measurement with window: the yield PE value is zero as expected.

---
[14] Lower values (fractions of the film thickness) were not used as they were not modeled in the calculation of the effective absorption.

### 4.4.2 Effective photon yield in the sample cell

The *VUV-only* measurement of the absorber can be used to quantify the effective VUV photon yield ($PY_{eff}$) from an α-decay in the sample cell. Although this is extracted from the simulation framework, we describe it analytically in this section in order to discuss its degeneracies with other optical parameters and uncertainties.

$PY_{eff}$ is related to the number of photons detected by the PMT ($N_{det}$), according to the equation:

$$N_{det} = PY_{eff} \cdot QE_{PMT}^{VUV} \cdot \Omega \qquad (6)$$

where $QE_{PMT}^{VUV}$ is the QE of the PMT at 128 nm and $\Omega$ is the fraction of solid angle covered by the PMT in the hemisphere above the source disk. A schematic of $\Omega$ in the absorber measurement is shown in Fig. 14 (left).

$PY_{eff}$ is also a function of the photon yield by an α depositing $E = 5.486$ MeV in LAr ($PY_{LAr}^\alpha = E \cdot SY_{LAr}$), but takes into account that part of the scintillation photons may be reflected by the source disk. That is, for the isotropic emission of photons around the α-decay, half of the photons can directly propagate through the cell, and the other half reach the surface of the source disc: being either absorbed or reflected by it, with a probability given by the reflectivity of the source disc, $SD_{ref}$. $PY_{eff}$ is thus defined as:

$$PY_{eff} = PY_{LAr}^{\alpha} \cdot 1/2 \cdot (1 + SD_{ref}) \quad (7)$$

And in terms of the scintillation yield ($SY_{LAr}$), we have:

$$PY_{eff} = E \cdot SY_{LAr} \cdot 1/2 \cdot (1 + SD_{ref}) \quad (8)$$

Equation 8 shows that $SY_{LAr}$ and $SD_{ref}$ are degenerate components of $PY_{eff}$. This means that their specific values are not important, as long as $PY_{eff}$ is well known. However, in case the reflectivity of the source disk, $SD_{ref}$, presents a specular component, $PY_{eff}$ is not isotropic. To investigate this uncertainty, we set $SD_{ref}$, as given in Sect. 4.4.1, and simulate the cases for different reflection components of $SD_{ref}$: part lambertian and the other part as 25, 40 or 60% specular.

This introduces a systematic uncertainty in determining $PY_{eff}$ from the absorber measurement. This uncertainty adds to the one from the measured PE value of the absorber and yields a total systematic uncertainty of $\pm 5\%$ on $PY_{eff}$. As the impurity level (and scintillation yield) do not vary significantly across measurements (as discussed in Sect. 4.3), uncertainties from varying $SY_{LAr}$ due to impurities are very likely covered by the $\pm 5\%$ uncertainty on $PY_{eff}$, and no further systematic uncertainty is associated to it.

Equation 6 shows that different values of $PY_{eff}$ and $QE_{PMT}^{VUV}$ can result in the measured $N_{det}$ for the given $\Omega$. We chose to fix $QE_{PMT}^{VUV}$ to the value given in Sect. 4.4.1, and fit the last free parameter of $PY_{eff}$: $SY_{LAr}$. The value for $SY_{LAr}$, so that $N_{det}$ results in the measured PE value, is $\sim 25 ph/keV$. We note that fixing $QE_{PMT}^{VUV}$ to estimate $PY_{eff}$ does not introduce any uncertainty on the estimation of efficiencies of other WLS samples as long as the ratio $q_{QE}$ between the true $QE_{PMT}^{vis}$ and $QE_{PMT}^{VUV}$ is equal to the that input in the simulation – that is, in the case the overall QE of the PMT increases in LAr (as discussed in Sect. 4.4.1), the only requirement is that its relative value remains the same. This is described by the relation:

$$QE_{PMT}^{vis} = q_{QE} \cdot QE_{PMT}^{VUV} \quad (9)$$

In this case,[15] the number of scintillation photons detected from any sample can be described similarly to the Eq. 6 from the absorber:

$$N_{det} = PY_{eff} \cdot QE_{PMT}^{VUV} \cdot (\Omega + \epsilon \cdot q_{QE}^s) \quad (10)$$

where $\epsilon$ is the sample enhancement factor: more photons reach the PMT due to shifting and/or reflection from the sample (as shown in Fig. 14, right) – and $q_{QE}^s$ is equal to $q_{QE}$ for the *VUV+vis* measurements or equal to unity for the *VUV-only* measurements. The extra light yield can be thus used to determine the QE or the VUV reflectivity of the samples. As these parameters involve the geometry and optics of the samples, we extract these values from the simulations, as described in the next subsections.

---
[15] Uncertainties from a different case are further considered in Sect. 4.4.4.

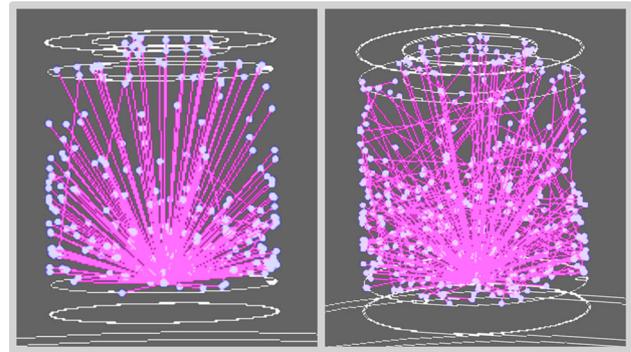

**Fig. 15** Left: simulation of an $\alpha$-decay in the LAr setup covered by the absorber. The $\alpha$ particle cannot be seen, only the VUV photons (in a reduced number, for clarity). Right: the sample surface in the simulation is set to Tetratex and a few photons are now reflected by its surface

*4.4.3 VUV-reflectivity of TTX*

To extract the VUV reflectivity of TTX, we simulated the TTX sample in the setup for different values of its VUV reflectivity. An example of the VUV photons being reflected by the TTX in the simulation is shown in Fig. 15.

We then find the value of reflectivity that yields the measured PE value from TTX. We run the simulation for the mean and limit ($\pm 5\%$ uncertainty) values of $PY_{eff}$ – this adds a systematic uncertainty to the result. The uncertainty from the measured PE value of the sample leads to a statistical uncertainty. The resulting VUV reflectivity is: $9 \pm 6(stat) \pm 2(syst)\%$. This means that although the measured PE value from TTX is larger than that of the absorber, we cannot significantly claim a non-zero reflectivity from it. This is consistent with the fact that their measured PE values agree within uncertainties. We thus set an upper limit (at 90% CL) that its VUV reflectivity is below 17%.

*4.4.4 QE of TPB and PEN*

The estimation of the QE of TPB and PEN is done in a similar way. The simulation is set for the given geometry and optical parameters, and then run for different values of QE of the shifters until we find the QE that yields the measured PE value. The difference for the WLSR samples is that, apart from the systematical uncertainty on $PY_{eff}$, we introduce further sample-related uncertainties in the simulations. These are discussed in detail in Sect. 4.4.1 and the main ones are listed in Table 4.

**Table 4** Range of expected and limit values of the optical parameters of the WLSR samples given their uncertainties. The limit values used for the *low limit case* of QE are shown in bold. See text (Sect. 4.4.4) for details

| TPB | | | |
|---|---|---|---|
| Parameter | Lower value | Expected value | Upper value |
| VUV-$\lambda_{abs}$ | **250 nm** | 350 nm | 450 nm |
| vis-$\lambda_{abs}$ | $M_v - \delta$ | $M_v$ | $\mathbf{M_v + \delta}$ |
| VUV-ref. TTX | 0% | 9% | **17%** |
| vis-ref. TTX | Unconstrained | 95% | **95%** |
| PEN | | | |
| Parameter | Lower value | Expected value | Upper value |
| vis-$\lambda_{abs}$ | $M_v - \delta$ | $M_v$ | $\mathbf{M_v + \delta}$ |
| vis-ref TTX | Unconstrained | 95% | **95%** |

For the TPB-based sample, we consider the uncertainties on the VUV and vis-$\lambda_{abs}$, as well as on the VUV to vis-reflectivity of TTX. For the PEN-based one, these VUV parameters are not relevant as we assumed that all VUV light is absorbed. For both samples, we consider the mean value ($M_v$) of vis-$\lambda_{abs}$ and its $\pm \delta$ uncertainty (described in Sect. 4.4.1). The uncertainties on the VUV-reflectivity of TTX in LAr were discussed in the previous section. For its vis-reflectivity, we consider that its value could be lower in LAr (in comparison to ≈ 95% measured in air) and thus do not place a limit on its lower value (*unconstrained*). By lowering the vis-reflectivity of TTX, the estimated QE of TPB increases and goes well above unity for vis-reflectivity < 80%. This means that the vis-reflectivity of TTX in LAr[16] is likely larger than 80%, but it is not possible to confidently constrain the upper value of the QE of TPB in LAr.

We thus focus on the *expected case* – which uses the most likely values of the optical parameters – and a *low limit case* – which considers the lower or upper values of these parameters (shown in bold in Table 4). For the latter, the estimation of the QE results in the lowest (most conservative) value. We thus use the *low limit case* to obtain 90% CL lower limits on the QE of TPB and PEN, which are shown in Table 5.

In the same table, we also show the results for the *expected case*, which considers only the uncertainties on the measured PE value and on $PY_{eff}$. The lower limit takes additionally all the uncertainties on the optical parameters of the shifters and reflector into account. It also covers other types of uncertainties that would increase the estimated QE if introduced in the simulation, such as the possible larger enhancement of the QE of the PMT in the VUV range (discussed in Sect. 4.4.2), the uncertainties on the emission spectrum of PEN, and on the VUV refractive indeces of TPB and PEN (discussed in

Sect. 4.4.1). We thus consider these as conservative limits. All values were extracted using the measurements and simulations of the *VUV+vis* mode.

For the *vis-only* mode, the simulation of the setup was run with the acrylic window and the expected values obtained for the QE of TPB and PEN were: $91 \pm 4(stat) \pm 4(syst)\%$ and $57 \pm 3(stat) \pm 4(syst)\%$, respectively. These values agree within errors with the expected values shown in Table 5. The *vis-only* estimations are however only a cross-check of the stability and of the shifted light. They are not further considered in the estimation of the lower limits since the acrylic window introduces further optical uncertainties, especially for PEN, which presents some uncertainty on its emission spectrum (as discussed in Sect. 4.4.1).

### 4.5 Discussion and conclusion

TTX, PEN and TPB were measured in a LAr setup with a VUV-vis sensitive PMT. The optical parameters of the shifters, reflector and of the setup were characterized and served as input in a simulation used to extract the VUV reflectance of TTX and the QE of the TPB and PEN.

The performance of the thin TTX film in reflecting LAr scintillation was well below that of other PTFE materials in response to scintillation from liquid xenon: < 17% compared to > 95% [55]. The observed low reflectance could be, for instance, due to the absorption edge of PTFE at ∼ 160 nm [56]. By measuring TTX coupled to shifters, we demonstrated that its performance in reflecting shifted light in LAr is better than 80% (and could be as good as in air, ∼ 95%).

An extensive characterization of the samples was performed with microscopy (Appendix A). in order to ensure their quality for the QE estimation of this work. The lower limit (67%) and expected values of the QE of TPB are consistent with the value from [20] corrected for the $21 \pm 13\%$ increase in efficiency observed by [22] at 87 K: $73 \pm 9\%$. While the measurements from this work result in an expected

---
[16] One surface of TTX faces LAr and the other surface boundary is TPB, but most of TTX is rather in LAr, since it is a porous material with thickness much larger than the TPB coating.

**Table 5** QE of TPB and PEN using the expected values of the optical parameters and 90% CL lower limit on the QE of TPB and PEN taking all investigated uncertainties into account

| QE | *Expected* value | 90% CL lower limit |
|---|---|---|
| TPB | $85 \pm 5(stat) \pm 6(syst)\%$ | 67% |
| PEN | $69 \pm 4(stat) \pm 5(syst)\%$ | 49% |

QE value of TPB lower than unity, they do not set an upper limit on it.[17]

The efficiency of PEN had been previously reported only relative to TPB. Although our results are independent of TPB, we can compare the light yield (LY) of PEN to that of TPB, as both were measured in the same setup. For this, we use the LY enhancement measured from PEN, relative to that of TPB: $E_{LY}^{rel} = E_{LY}^{PEN}/E_{LY}^{TPB}$[18]. This ratio and relative LY values from the literature are listed in Table 6.

These LY ratios are however not directly comparable: while the QE is an intrinsic property of the material, the measured LY depends on the geometry of the setup and optics of the sample. For instance, the photodetector coverage in the setup of [28] was ∼1%: photons thus scattered many times on the samples, and the relative LY was thus more dependent on their optical parameters such as reflectivity or absorption. In the setup of this work, the photodetector coverage was ∼6% and photons were less likely to scatter multiple times because the bottom was covered with the absorber.

As the LY depends on the optical parameters of the sample, it varies for different substrates, coating thickness and method. Large variations of LY were measured from TPB with different thickness by [13,20,21,29] and PEN from different thickness/types by [27]. This large sample dependence can lead to large differences among results of relative efficiency of LY, as observed in Table 6. The highest LY from PEN relative to TPB measured directly in LAr is the value from this work[19].

It is however unclear whether this large relative LY is also related to a better performance of our specific PEN-based WLSR, such as a better efficiency of TTX as a reflector for PEN in LAr.[20] Another possibility is that sanding PEN increased its LY by allowing light to more easily leave the film (as shown in Fig. 20): photons seem to be less prone to total internal reflection and subsequent absorption. This increase in LY from sanded PEN was observed in [32] and [15], being regarded by the latter as a hint that sanding removes a possibly degraded layer of the film.

The *expected* QE values estimated in this work indicate that the intrinsic QE of PEN may be close to that of TPB: $69 \pm 4(stat) \pm 5(syst)\%$ compared to $85 \pm 5(stat) \pm 6(syst)\%$, with a relative mean value of ∼ 81%. This value is higher than the relative LY values of Table 6. This indicates that the LY measured from a PEN-based WLSR can be close to that of a TPB-based one if a few of its optical parameters are improved, such as its absorption and coupling to the reflector (discussed in Sect. 3). While a few applications may require the efficiency of PEN to be very close to that from TPB, for other applications the lower limit on the QE of PEN (49%) measured in this work may be enough to approve its suitability.

## 5 Summary and outlook

The goal of this work was to develop and characterize an efficient WLSR for the LAr instrumentation of LEGEND-200 and for future LAr-based detectors. In this context, several optical parameters of TPB, PEN and TTX were measured.

The reflector TTX was measured in air and in LAr. While TTX reflects well visible light (∼ 95% measured in air, and likely over 80% in LAr), it reflects < 17% of scintillation light from LAr. This motivates the use of shifters not only to facilitate the detection of LAr's VUV scintillation, but also its reflection. TPB evaporated TTX reflected visible light at a similar intensity of bare TTX. When PEN is coupled to TTX, the intensity of reflection is lower: some photons are likely absorbed while traversing the PEN film or after several reflections (inside the film or between the film and TTX). These effects were observed with spectrophotometers and with a fluorescence microscope.

Given its enhanced performance, TPB on TTX was chosen for the WLSR of LEGEND-200. Its coating was proved to be homogeneous and its in-situ TPB vacuum evaporation was likely the largest one to-date. Future experiments may however require much larger surfaces covered with WLSR and such a TPB evaporation might be unpractical. Thus, PEN has been extensively investigated as an easy-to-scale alternative [27,28,58]. PEN's efficiency has been, however, always measured relative to TPB [18,27,28,32,58]. Here, we pro-vide for the first time measurements independent of TPB. The lower limits on the QE of PEN and TPB in LAr are 49 and 67%, respectively (at 90% CL).

---

[17] Efficiency values larger than unity have been reported for TPB [25,57]: the value from [57] has been later corrected [20], and the value from [25] was obtained with a different method which might not be directly comparable to the results from this work.

[18] To calculate $E_{LY}^{rel}$, we consider the PE value measured from PEN and TPB in the *VUV+vis* measurements subtracted by the one measured from the absorber (listed in Table 3).

[19] This higher relative value holds even for the PE values measured from PEN and TPB in the *vis-only* mode: which is > 48% despite the fact that the acrylic window cuts more from the emission from PEN than from TPB (an effect that is only modeled in the simulation).

[20] We note that TTX was air-coupled to PEN and that [27] observed that optically coupling the ESR reflector to PEN decreased its LY.

**Table 6** LY from PEN relative to TPB in response to 128 nm (LAr scintillation) light, at room temperature (RT) or LAr temperature ($\sim 87$ K). All the films are biaxially oriented: a few of them backed by a reflector, and a few sanded (marked with $(s)$). Geometry of the setups vary. Values from this work are shown in bold

| PEN sample | Relative LY | Reflector | Temperature |
| --- | --- | --- | --- |
| Molded PEN | $\sim 50\%$ [18] | No | RT |
| Film (125 µm)$^{(s)}$ | $\sim 50\%$ [32] | No | RT |
| Film (125 µm) | 80 (23)% [15] | No | LAr[a] |
| Film (125 µm) | 34 (1)% [27] | ESR | LAr |
| Film (25 µm) | 39 (2)% [28] | ESR | LAr |
| Film (125 µm)$^{(s)}$ | **75 (7)%** | TTX | LAr |

[a] The value from [15] corresponds to the relative efficiency measured at RT and projected at LAr temperature

The results from this work indicate that the QE of PEN in LAr is lower than that of TPB but not as low as previously reported values of relative LY. By measuring the parameters of PEN and TTX in detail, we show that there is room for improving the light yield from a PEN-based WLSR: while PEN is usually acquired in the form of films with specular surfaces, the rough surface of the PEN with TTX backing measured in this work yield more light than other specular PEN films backed with ESR [27,28]. Also, thinner PEN films could be used so that less light is lost due to absorption within the film.

The characterization of TPB and PEN performed in this work is informative for WLS selection in future detectors, and for the optical simulations of not only the LAr instrumentation of LEGEND-200 and the future experiment LEGEND-1000, but also of other experiments that use TPB or PEN. As the QE of shifters is often degenerate with other optical parameters of detectors, constraining their intrinsic value is of key importance to constrain other parameters in the simulations of detectors that use these WLS films.

The limiting factors in constraining the QE of the shifters were the uncertainties on the VUV-absorption length of TPB and on the visible reflectivity of TTX in LAr.

**Acknowledgements** We would like to thank Marcin Kuźniak for fruitful discussions and comments. We also thank Patricia Sanchez and Michelle Galloway for proof-reading the manuscript and for comments. We gratefully acknowledge support from the Swiss National Science Foundation under Grants no. 200020-188716 and 20FL20-201537, from the University of Zurich under the Candoc Grant no. K-72312-09-01, and from the European Research Council (ERC) under the European Union's Horizon 2020 research and innovation programme, Grant agreement No. 742789 (Xenoscope). We are grateful to the Center for Microscopy and Image Analysis of the University of Zurich for the microscopy imaging, which was performed with their equipment and support.

**Data Availability Statement** This manuscript has no associated data or the data will not be deposited. [Authors' comment: The data supporting the results of this work is available from the corresponding author, upon reasonable request.]

## Appendix A: Microscopy imaging

To investigate whether the in-situ evaporated TTX of LEGEND-200 was homogeneously covered with TPB, we imaged its 'witness' sample with microscopes. This characterization is relevant not only as a check of the evaporation process but also to qualify the sample for the QE estimation done in this work: if the sample was not uniformly covered with TPB, its QE would be under-estimated.

As a single TPB molecule is much smaller than the pores of TTX (1 nm[21] compared to 250 nm [59]), we use scanning electron microscopy (SEM) to check whether TPB fills the pores in TTX. SEM is also used to characterize the molecule ordering and the crystalline structure of TPB in the sample, which can take different forms, depending on the coating technique and cooling rate [60–62]. To check the homogeneity of the coating at larger ranges, we scan larger samples with fluorescence microscopy.

We also image the PEN film, in order to investigate its sanded surface and how the photons emitted by it reflect within the film.

---

[21] Value estimated by the sum of benzene rings and C-C bonds, or by dividing its molar mass (358.48 g/mol) by its density (1.1 g cm$^{-3}$).

Appendix A.1: Procedures and setups

All the samples prepared for the tests in LAr were measured. To investigate the effects of LAr immersion and ambient light exposure on TTX+TPB (L), one sample of it went through 2 runs of a five-day-long immersion in LAr, and one sample of it was partially exposed to ambient light for about two weeks.[22] As a comparison of the quality of the TTX+TPB (L) coating achieved in the in-situ evaporation, a sample of TTX dip-coated with TPB (as described in [29]) was measured. A list of the sample names and their description is given in Table 7.

For the SEM images, the samples were sputter coated with 2 nm of platinum using a Safematic sputer coater.[23] Each sample was scanned in 3–4 spots with a Zeiss GeminiSEM 450 field emission SEM. Image sizes varied from around 1 mm to 1 μm and resolution of $\mathcal{O}(10\,\text{nm})$.

For the fluorescence microscopy scans, a Widefield Leica Thunder Imager 3D was used. The wavelength of the excitation light was 390 nm and an additional filter of range between 375 nm to 405 nm ensured the light was monochromatic. A dichroic mirror with a cutoff at 415 nm prevented the 390 nm light from reaching the detectors in case it was reflected by the sample. The fluorescence response from the sample (with wavelength above 415 nm) was collected by the objective and recorded by a camera.

Appendix A.2: Results and discussion

SEM images of the bare TTX sample are shown in the left image of Fig. 16, and in the top images of Fig. 17. The first figure shows that the tubular structures of TTX have a diameter close to that measured by [59] (250 nm), and that the apparently large pores are are well covered with TPB. The images in Fig. 17 show that the TPB coverage and distribution is homogeneous also at larger scales (of a few micrometers): no parts clearly lack on TPB.[24]

The crystalline shape of TPB shown in these figures is needle-like, which is the common shape of the alpha form of TPB [61]. Out of the four structure types that have been reported for TPB [60–62], the alpha is the most frequent and most stable form at room temperature [61]. It is also the form that produces a tighter packing, with a UV-vis absorption spectrum different from the one from other crystalline forms, such as the beta [62]. Note that the needle structures are several times larger than the nominal coated thickness, as also observed by [63] and [64].

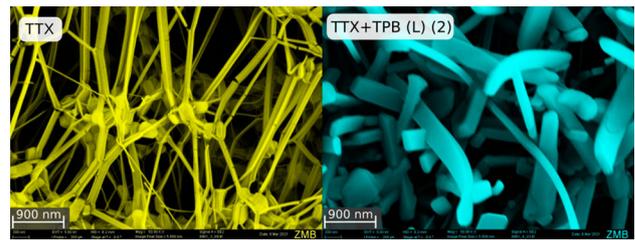

**Fig. 16** Close (300 nm scale bar) SEM images of TTX and TTX+TPB (L) (2). The images were colored in yellow and blue, respectively, to ease the observation of their structures

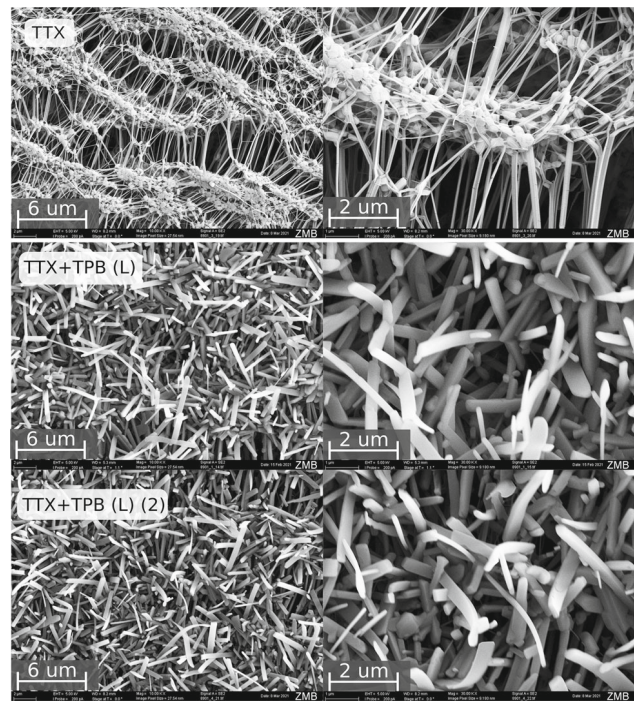

**Fig. 17** SEM images of TTX, and TTX+TPB (L) before and after (2) immersion in LAr. The scale bars of the figures on the left and right are 6 and 2 μm respectively

---

[22] The sample was kept inside a glass box and half of it was covered – so that half of the sample was exposed to light – but none of it was directly exposed to humidity and dust.

[23] The Pt coating avoids charge collection by the samples and stabilizes them mechanically so that the high-energy electrons do not damage their surface.

[24] Not all structures are visible in an SEM image: the SEM detector receives less scattered electrons from areas with less or none Pt coating.

Even though this crystal form is dominant in films produced by sublimation [60,61], SEM images from [51] show that vacuum evaporated TPB samples not always present "needles" nor are always homogeneous: the form and homogeneity do not depend solely on the coating technique. Even for the same coating technique, substrate and TPB film thicknesses, different evaporation conditions can result in variations of the light yield [34].

The third row of images in Fig. 17 shows the TTX+TPB (L) after 2 immersions in LAr. Similarly to reference [63],

**Table 7** Exposure and description of the samples measured with microscopes. Samples marked with (*) were described in Sect. 2

| Name | Description | Measurement |
| --- | --- | --- |
| TTX | Bare sample∗ | SEM |
| TTX+TPB (L) | Evaporation witness sample* kept in dark | SEM and fluorescence mic |
| TTX+TPB (L)(2) | TTX+TPB (L) after 2 immersions in LAr | SEM |
| TTX+TPB (L)(E) | TTX+TPB (L) partially exposed to ambient light | Fluorescence microscopy |
| TTX+TPB (D) | TPB dip-coated kept in dark | Fluorescence microscopy |
| PEN | Bare sanded film∗ | Fluorescence mic |

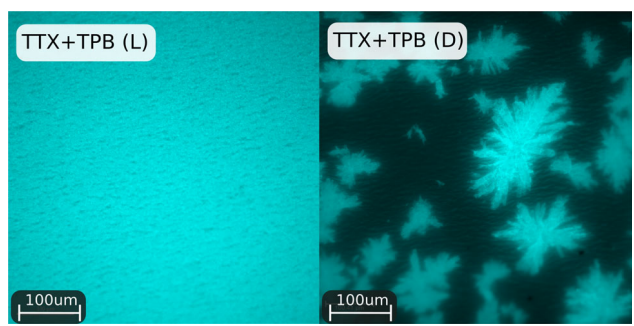

**Fig. 18** Fluorescence microscopy images of TTX+TPB (L) and TTX+TPB (dip-coated). The scale bars are 100 μm

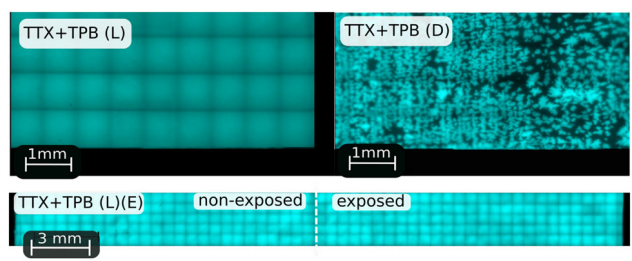

**Fig. 19** Top: fluorescence microscopy images of TTX+TPB (L) and TTX+TPB (dip-coated). Images are composed of multiple scans ('tiles'). The scale bars are 1 mm. Bottom: A long tile scan of TTX+TPB (L). The left half of it had been exposed to ambient light for two weeks. A shaded line in the middle identifies the border between exposed and non-exposed sample. The scale bar is 3 mm

we do not see any damage of the needle-like structures of TPB after immersion in LAr.[25]

The left figures of Figs. 18 and 19 show the single and multiple fluorescence microscopy scans of TTX+TPB (L). The scans of dip-coated TTX+TPB are shown on the right for comparison: these show several large crystals and less light from the parts without crystals. Quantifying the QE of TPB from such a sample would lead to large errors, given that large parts of it are not properly covered with TPB. The TTX+TPB (L) sample, on the contrary, is very homogeneous: its light does not go to zero (or close to it) at any point. The tile scans only present a shadow on the edges, which is an effect from the central focusing point.

The bottom image of Fig. 19, shows that the moderate (2 week long) exposure of one side of the sample to ambient light did not decrease the intensity of its fluorescence in response to UV (390 nm) light.

Figure 20 shows that sanding PEN produced scratches with a few micrometers width. Also noticeable is that more light is observed from the scratches and edges of the film, indicating that the photons emitted by PEN may get "trapped" inside the film due to total internal reflection.[26] This is expected to happen also in LAr, since the refractive index

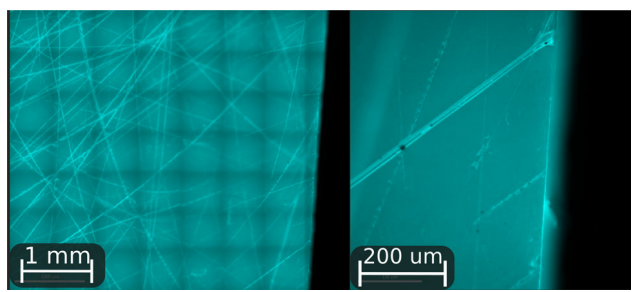

**Fig. 20** Fluorescence microscopy images of PEN. A tile scan is shown on the left and a single scan on the right. The scale bars are 1 mm and 200 μm respectively

of LAr at visible wavelengths is not much higher than that of air [43] and still much lower than that of PEN [53].

### References

1. DarkSide Collaboration, P. Agnes, et al., DarkSide-50 532-day dark matter search with low-radioactivity argon. Phys. Rev. D **98**(10), 102006 (2018). arXiv:1802.07198
2. DEAP Collaboration, P.A. Amaudruz, et al., Design and construction of the DEAP-3600 dark matter detector. Astropart. Phys. **108**, 1–23 (2019). arXiv:1712.01982
3. DUNE Collaboration, B. Abi, et al., The single-phase ProtoDUNE technical design report. (2017). arXiv:1706.07081
4. J.-J. Wang, MiniCLEAN dark matter experiment. PhD thesis, The University of New Mexico (2017)

---

[25] This however does not quantify whether there is a slow diffusion of TPB into LAr, as observed by [19].

[26] While the probability of total internal reflection of light in the vis measurements was not high (the angle of incidence was 8°), the probability of total internal reflection by photons emitted isotropically inside the film is much larger.

5. MicroBooNE Collaboration, R. Acciarri, et al., Design and construction of the MicroBooNE detector. JINST **12**(02), P02017–P02017 (2017)
6. ArDM Collaboration, J. Calvo, et al., Commissioning of the ArDM experiment at the Canfranc underground laboratory: first steps towards a tonne-scale liquid argon time projection chamber for dark matter searches. J. Cosmol. Astropart. Phys. **2017**(03), 003–003 (2017)
7. GERDA Collaboration, M. Agostini et al., Upgrade for Phase II of the Gerda experiment. Eur. Phys. J. C **78**(5), 388 (2018). arXiv:1711.01452
8. D. Akimov, et al., COHERENT 2018 at the spallation neutron source. (2018). arXiv:1803.09183
9. LEGEND Collaboration, et al., LEGEND-1000 preconceptual design report (2021)
10. M. Hofmann, et al., Ion-beam excitation of liquid argon. Eur. Phys. J. C **73**(10), 2618 (2013), arXiv:1511.07721
11. G.R. Araujo, T. Pollmann, A. Ulrich, Photoluminescence response of acrylic (PMMA) and polytetrafluoroethylene (PTFE) to ultraviolet light. Eur. Phys. J. C **79**(8), 653 (2019). arXiv:1905.03044
12. J. Kratz, Investigation of tetraphenyl butadiene coatings for wavelength shifting fibers for the LAr veto in GERDA. Master's thesis, Technical University of Munich (2016)
13. M. Schwarz, et al., Liquid argon instrumentation and monitoring in LEGEND-200. EPJ Web Conf. ANIMMA 2021 - Advancements in Nuclear Instrumentation Measurement Methods and their Applications, vol. 253 (2021)
14. GERDA Collaboration, M. Agostini et al., Final results of GERDA on the Search for Neutrinoless Double-$\beta$ decay. Phys. Rev. Lett. **125**(25), 252502 (2020). arXiv:2009.06079
15. M. Kuźniak et al., Polyethylene naphtalate film as a wavelength shifter in liquid argon detectors. Eur. Phys. J. C **79**(4), 291 (2019). arXiv:1806.04020
16. M. Kuźniak, A.M. Szelc, Wavelength shifters for applications in liquid argon detectors. Instruments **5**(1), (2021). https://doi.org/10.3390/instruments5010004
17. Okimichi Yano, Hitoshi Yamaoka, Cryogenic properties of polymers. Prog. Polym. Sci. **20**(4), 585–613 (1995)
18. Y. Efremenko et al., Use of poly(ethylene naphthalate) as a self-vetoing structural material. JINST **14**(07), P07006–P07006 (2019)
19. J. Asaadi et al., Emanation and bulk fluorescence in liquid argon from tetraphenyl butadiene wavelength shifting coatings. JINST **14**(2), P02021 (2019). arXiv:1804.00011
20. C. Benson, G.O. Gann, V. Gehman, Measurements of the intrinsic quantum efficiency and absorption length of tetraphenyl butadiene thin films in the vacuum ultraviolet regime. Eur. Phys. J. C **78**(4), 329 (2018). arXiv:1709.05002
21. C.H. Lally et al., UV quantum efficiencies of organic fluors. Nucl. Instr. Methods Phys. B **117**(4), 421–427 (1996)
22. R. Francini et al., VUV-Vis optical characterization of tetraphenyl-butadiene films on glass and specular reflector substrates from room to liquid Argon temperature. JINST (2013). arXiv:1304.6117
23. C.S. Chiu et al., Environmental effects on TPB wavelength-shifting coatings. JINST **7**(07), P07007–P07007 (2012). arXiv:1204.5762
24. Stacie E. Wallace-Williams et al., Excited state spectra and dynamics of phenyl-substituted butadienes. J. Phys. Chem **98**(1), 60–67 (1994)
25. Joshua R. Graybill et al., Extreme ultraviolet photon conversion efficiency of tetraphenyl butadiene. Appl. Opt. **59**(4), 1217–1224 (2020)
26. J.M. Corning et al., Temperature-dependent fluorescence emission spectra of acrylic (PMMA) and tetraphenyl butadiene (TPB) excited with UV light. JINST **15**(03), C03046–C03046 (2020)
27. Y. Abraham et al., Wavelength-shifting performance of polyethylene naphthalate films in a liquid argon environment. JINST **16**(07), P07017 (2021)
28. M.G. Boulay, et al., Direct comparison of PEN and TPB wavelength shifters in a liquid argon detector. (2021). arXiv:2106.15506
29. L. Baudis et al., Enhancement of light yield and stability of radio-pure tetraphenyl-butadiene based coatings for VUV light detection in cryogenic environments. JINST **10**(09), P09009–P09009 (2015)
30. J.R. Lacowicz, *Principles of Fluorescence Spectroscopy*, 3rd edn. (Springer, Berlin, 2006)
31. Donaldson.Website
32. G.R. Araujo. Wavelength shifting and photon detection of scintillation light from liquid argon. Master's thesis, Tech. Univ. Munich (2019)
33. Goodfellow.Website
34. A. Leonhardt, Characterization of wavelength shifters for rare event search experiments with a VUV spectrofluorometer. Master's thesis, TUM (2021)
35. Data sheet of the Metal Velvet™ foil from Acktar
36. Data sheet of the Hamamatsu. R955 PMT
37. D. Arun Kumar et al., Growth and characterization of organic scintillation single crystal 1,1,4,4-tetraphenyl-1,3-butadiene (tpb) using vertical bridgman technique. Optic. Mater. **109**, 110286 (2020)
38. M. Walter, Background reduction techniques for the Gerda experiment. PhD thesis, Universität Zürich (2015)
39. V.H.S. Wu, Low energy calibration for GERDA and characterization of wavelength-shifters and reflectors in liquid argon for LEGEND-200. Master's thesis, University of Zurich (2021)
40. WArP Collaboration, R. Acciarri et al., Effects of nitrogen contamination in liquid argon. JINST **5**, P06003 (2010). arXiv:0804.1217
41. DEAP Collaboration, P. Adhikari et al., The liquid-argon scintillation pulseshape in DEAP-3600. Eur. Phys. J. C **80**(4), 303 (2020). arXiv:2001.09855
42. CERN: Geant4 toolkit.Website
43. M. Babicz et al., A measurement of the group velocity of scintillation light in liquid argon. JINST **15**(09), P09009–P09009 (2020)
44. A. Neumeier et al., Attenuation of vacuum ultraviolet light in pure and xenon-doped liquid argon—an approach to an assignment of the near-infrared emission from the mixture. EPL **111**(1), 12001 (2015)
45. ArDM Collaboration, J. Calvo et al., Measurement of the attenuation length of argon scintillation light in the ArDM LAr TPC. Astropart. Phys. **97**, 186–196 (2018). arXiv:1611.02481
46. N. Ishida et al., Attenuation length measurements of scintillation light in liquid rare gases and their mixtures using an improved reflection suppressor. Nucl. Instrum. Methods A **384**(2), 380–386 (1997)
47. Data sheet for the R11065 modified PMT – data provided by Hamamatsu
48. H.H. Li, Refractive index of alkaline earth halides and its wavelength and temperature derivatives. J. Phys. Chem. Ref. Data **9**(1), 161–290 (1980)
49. P. Laporte et al., Vacuum-ultraviolet refractive index of LIF and MGF 2 in the temperature range 80–300 k. J. Opt. Soc. Am. **73**, 1062–1069 (1983)
50. B. Zatschler, Attenuation of the scintillation light in liquid argon and investigation of the double beta decay of $^{76}$Ge into excited states of $^{76}$Se in the GERDA experiment. PhD thesis, Dresden, Tech. U. (2020)
51. D. Stolp et al., An estimation of photon scattering length in tetraphenyl-butadiene. JINST **11**(03), C03025–C03025 (2016)
52. D. Mary et al., Understanding optical emissions from electrically stressed insulating polymers: electroluminescence in poly(ethylene terephthalate) and poly(ethylene 2,6-naphthalate) films. J. Phys. D Appl. Phys. **30**, 171 (1999)
53. N. Hong, R.A. Synowicki, J.N. Hilfiker, Mueller matrix characterization of flexible plastic substrates. Appl. Surf. Sci. **421**, 518–528 (2017). 7th Inter. Conference on Spectroscopic Ellipsometry

54. Xiaoning Zhang et al., Complex refractive indices measurements of polymers in visible and near-infrared bands. Appl. Opt. **59**(8), 2337–2344 (2020)
55. F. Neves et al., Measurement of the absolute reflectance of polytetrafluoroethylene (PTFE) immersed in liquid xenon. JINST **12**(01), P01017 (2017). arXiv:1612.07965
56. M.K. Yang, R.H. French, E.W. Tokarsky, Optical properties of Teflon® AF amorphous fluoropolymers. J. Micro/Nanolith. MEMS MOEMS **7**(3), 033010 (2008)
57. V.M. Gehman et al., Fluorescence efficiency and visible re-emission spectrum of tetraphenyl butadiene films at extreme ultraviolet wavelengths. Nucl. Instrum. Methods Phys. A **654**(1), 116–121 (2011)
58. M. Kuźniak et al., Development of very-thick transparent GEMs with wavelength-shifting capability for noble element TPCs. Eur. Phys. J. C **81**(7), 609 (2021). arXiv:2106.03773
59. S.C. Amouroux, D. Heider, J.W. Gillespie, Characterization of membranes used in pressure driven composite processing. Compos. Part A Appl. Sci. Manuf. **41**(2), 207–214 (2010)
60. Alessia Bacchi et al., Raman investigation of polymorphism in 1,1,4,4-tetraphenyl-butadiene. J. Raman Spectrosc. **44**(6), 905–908 (2013)
61. A. Bacchi et al., Exploration of the polymorph landscape for 1,1,4,4-tetraphenyl-1,3-butadiene. CrystEngComm **16**, 8205–8213 (2014)
62. A. Girlando et. al., Spectroscopic and structural characterization of two polymorphs of 1,1,4,4-tetraphenyl-1,3-butadiene. Cryst. Growth Des. **10** (2010). https://doi.org/10.1021/cg100253k
63. B. Broerman et al., Application of the TPB wavelength shifter to the DEAP-3600 spherical acrylic vessel inner surface. JINST **12**, 04017 (2017). arXiv:1704.01882
64. Yikui Fan et al., Effects of alpha particle and gamma irradiation on the 1,1,4,4-tetraphenyl-1,3-butadiene wavelength shifter. Radiat. Phys. Chem. **176**, 109058 (2020)